\documentclass[final,numberedheadings]{aipproc}

\layoutstyle{6x9}

\def\lesssim{\mathrel{\hbox{\rlap{\hbox{\lower3pt\hbox{$\sim$}}}{\lower-1pt\hbox{$<$}}}}}
\def\gtrsim{\mathrel{\hbox{\rlap{\hbox{\lower3pt\hbox{$\sim$}}}{\lower-1pt\hbox{$>$}}}}}
\newcommand{\nablas}{{\bf{\nabla}}\!_s}

\begin{document}

\title[Large-scale magnetic fields in GRB outflows]
{Large-scale magnetic fields in GRB outflows:
acceleration,$\!$ collimation,$\!$ and neutron decoupling}

\author{Nektarios Vlahakis}{
address={Section of Astrophysics, Astronomy \& Mechanics,
Dept. of Physics, University of Athens, Greece}}
\author{Arieh K\"onigl}{
address={Dept. of Astronomy \& Astrophysics and Enrico Fermi Institute, 
University of Chicago}}

\begin{abstract}
Using ideal magnetohydrodynamics we examine an outflow from a
disk surrounding a stellar-mass compact object.
We demonstrate that the magnetic acceleration is efficient
($ \gtrsim 50\%$ of the magnetic energy
can be transformed into kinetic energy of $\gamma> 10^2$ baryons) and also
that the jet becomes collimated to very small opening angles.
Observational implications, focusing on the case of an initially 
neutron-rich outflow, are discussed in K\"onigl's contribution.
\end{abstract}

\maketitle

\section{Ideal Magnetohydrodynamics}
There is growing evidence in favor of magnetic driving in 
outflows associated with gamma-ray burst (GRB) sources 
\citep[e.g.,][hereafter VK03a; 
see also K\"onigl's contribution in these Proceedings]{VK03a}.
The dynamics of these outflows may be described to zeroth order 
by the ideal, axisymmetric, hydromagnetic equations, consisting of
the Maxwell and momentum equations together with the conservations of
baryonic mass and specific entropy.
VK03a demonstrated that,
under the assumptions of a quasi-steady poloidal magnetic field
and of a highly relativistic poloidal velocity,
these equations become effectively time-independent and the 
motion can be described as a frozen pulse, 
generalizing  the so-called ``frozen pulse'' approximation
already known in purely hydrodynamic models of GRB outflows \citep{PSN93}.
Introducing the magnetic flux function $A$,
the arclength along a poloidal streamline $\ell$,
and the operator $\nablas$ that acts while keeping $s \equiv ct - \ell$ constant,
the momentum equation can be written as (see VK03a for details)
\begin{equation}\label{mom}
\gamma \rho_0 
\displaystyle
\left({\bf{V}} \cdot \nablas \right) \left(\xi \gamma {\bf{V}} \right) 
=\frac{
\left( \nablas \cdot {\bf{E}} \right) {\bf{E}}
+ \left(\nablas
\times {\bf{B}} \right) \times {\bf{B}}
}{4 \pi} -{{\bf{\nabla}}P}
\,.
\end{equation}
The large-scale electromagnetic field (${\bf{E}}$, ${\bf{B}}$), the bulk flow speed (${\bf{V}}$), 
and the total ($e^\pm +$ radiation)
pressure can be written as functions of $A$ and the rest baryon density 
$\rho_0$\footnote{$(z\,, \varpi, \phi)$, and
$(r\,, \theta \,, \phi)$ denote cylindrical and spherical coordinates,
whereas subscripts $p$ and $\phi$ denote poloidal and azimuthal
components, respectively.} 
\begin{eqnarray}\label{eq1}
{\bf{B}}= \frac{\nablas A \times
\hat{\phi} }{\varpi} + {\bf{B}}_{\phi}\,, \quad
{\bf{E}}=-\frac{\Omega}{c} \nablas A\,, \quad
{\bf{V}}=\frac{A \Omega^2}{4 \pi \gamma \rho_0 c^3 \sigma_{\rm M} }
{\bf{B}} + \varpi \Omega \hat {\phi}\,, \quad
P=Q \rho_0^{4/3}\,.
\end{eqnarray}
Faraday's law and the conservations of specific entropy and mass imply that
the functions $\Omega$, $Q$, and $\sigma_{\rm M}$ are constants of motion,
i.e., functions of $A$.
By integrating equation (\ref{mom}) 
along ${\bf {V}}_p$ and $\hat{\phi}$ one gets
two additional constants of motion,
\begin{eqnarray}\label{eq2}
\mu c^2= \xi \gamma c^2 - \frac{c^3 \sigma_{\rm M} }{A \Omega} \varpi B_\phi \,, \quad
\frac{\mu c^2 x_{\rm A}^2 }{\Omega}
= \xi \gamma \varpi V_\phi -\frac{c^3 \sigma_{\rm M} }{A \Omega^2} \varpi B_\phi\,, 
\end{eqnarray}
describing the conservation of the ratio (total energy
flux)/(mass flux) and of the total specific angular momentum, respectively.
The remaining unknowns are the functions
$A(r\,, \theta)$ and $\rho_0(r\,, \theta)$.
The latter is the solution of Bernoulli's equation
[which is obtained after substituting all quantities in the identity
$\gamma^2=1+\gamma^2 V_p^2/c^2+\gamma^2 V_\phi^2/c^2$ using
eqs. (\ref{eq1}) and (\ref{eq2})], 
whereas the former controls the shape of the poloidal
streamlines and is the solution of the highly nonlinear
transfield component of the momentum equation (\ref{mom}).

VK03a integrated the last two equations under the $r$ self-similarity assumption
$A=r^F f(\theta)$, which makes it possible (with
$\Omega \propto A^{-1/F}$, $Q \propto A^{(4-2F)/3F}$, and
constant $F$, $\sigma_{\rm M}$, $\mu$, and $x_{\rm A}$)
to separate the ($r$, $\theta$)
coordinates. The resulting simplified set of ordinary
differential equations can be easily integrated. 
Two types of boundary conditions at the base of the flow were considered:
(1) The case of a strong poloidal magnetic field
$(B_p \gtrsim B_\phi)$, which corresponds to a trans-Alfv\'enic
outflow (since the azimuthal field dominates asymptotically and
the flow becomes super-fast). (2) The case of a strong azimuthal field $(B_p \ll B_\phi)$,
which corresponds to a super-Alfv\'enic flow.

\subsection{Trans-Alfv\'enic flows}
A representative solution is shown in Fig. \ref{fig1}. 
Looking at panel ($a$), which shows the acceleration, one can
distinguish three different regimes:
\begin{figure}
  \includegraphics[height=.565\textheight]{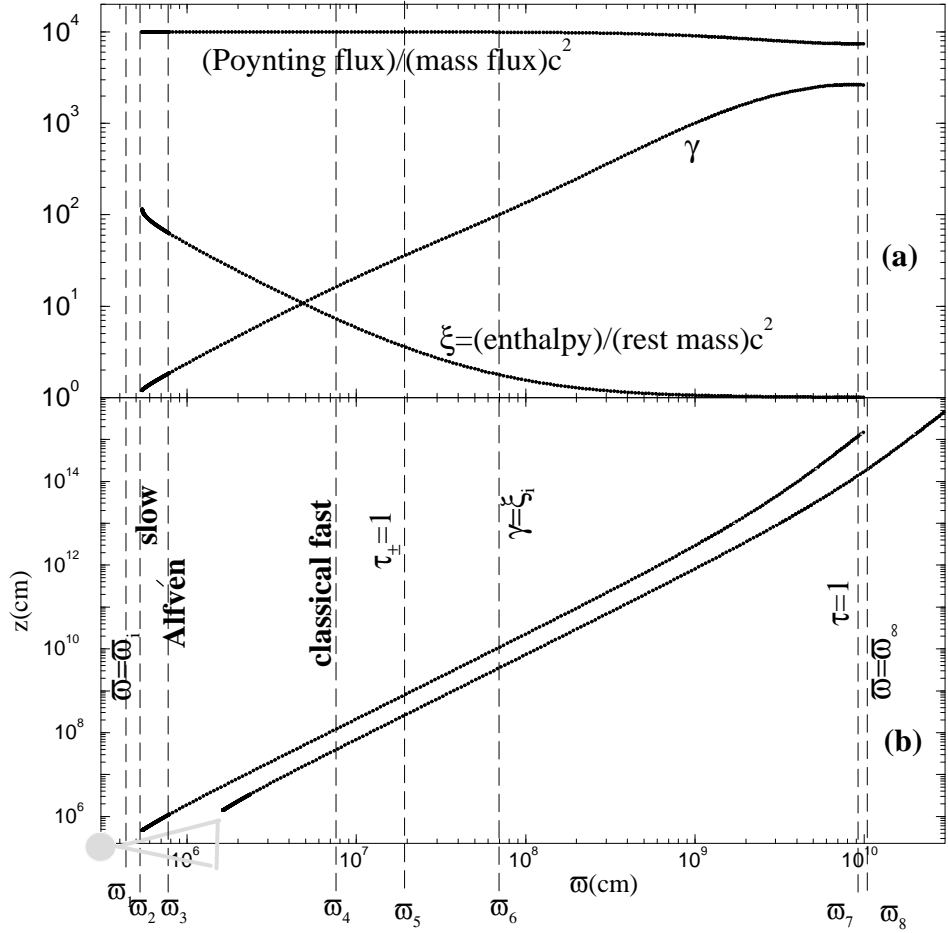}
\caption{
Trans-Alfv\'enic flow solution. ($a$) The Lorentz factor
$\gamma$, the ratio $\xi$ of the enthalpy to the rest energy, and the
ratio of the Poynting flux to the rest-energy flux ({\it top}\/ curve)
are shown as functions of $\varpi$
along the innermost field line. ($b$) The meridional projections
of the innermost and outermost field lines are shown on a
logarithmic scale, along with a sketch of the
central object/disk system.
The field lines have a parabolic shape ($z\propto \varpi^2$)
for $\varpi \lesssim 10^9$cm
and become asymptotically cylindrical.
The vertical lines mark the positions of the
various transition points along the
innermost field line \citep[see text and][]{VK01}.
}\label{fig1}
\end{figure}

\noindent 1)
$\varpi_1<\varpi<\varpi_6$ is the fireball phase.
The specific enthalpy $\xi$ decreases, resulting in increasing
$\gamma \propto \varpi$
($\xi \gamma \approx$ const, a characteristic of
hydrodynamic acceleration), while the specific Poynting flux remains constant
(the field is force free).
The electromagnetic field only guides the flow,
with the bulk of the collimation occurring in this regime.

\noindent 2)
$\varpi_6<\varpi<\varpi_8$ is the magnetic acceleration regime.
The fluid is cold ($\xi \approx 1$), but $\gamma$ continues to
increase (roughly as $\gamma \propto \varpi$) due to the decreasing
specific Poynting flux.

\noindent 3)
$\varpi=\varpi_8$ is the asymptotic cylindrical regime.
The final Lorentz factor is of the order of the final
specific Poynting flux, meaning that $\sim 1/2$ of the total energy
(which was mostly electromagnetic initially)
is transformed into baryonic kinetic energy ($\gamma_\infty \approx \mu /2$).

The solution presented in Fig. 1 describes one 
shell, corresponding to a particular value of $s$. By 
specifying the $s$ dependence in the initial conditions one can examine a 
multiple-shell outflow and the time dependence of the pulse. For
example, the $s$ dependence of the (total energy)/mass flux
ratio $\mu(s) c^2$ translates into
different final Lorentz factors for distinct shells:
$\gamma_\infty (s) \approx \mu(s)/2$.
In contrast with Michel's solution \citep{M69}, 
in which the classical fast magnetosonic point is located at
infinity and $\gamma_\infty \approx \mu^{1/3}$, here this
point is encountered at a finite height and most
of the magnetic acceleration occurs further out, leading
asymptotically to $\gamma_\infty(s) \approx \mu(s)/2 \gg \mu(s)^{1/3}$.
Thus, not only is the magnetic acceleration highly efficient,
but the stronger dependence of $\gamma_\infty(s)$ on the
initial conditions [through $\mu(s)$] can lead to a larger 
contrast in $\gamma_\infty$ between successive shells
and hence to a higher efficiency of internal shocks
\citep[e.g.,][]{P99}.

\subsection{Super-Alfv\'enic flows}
\begin{figure}
  \includegraphics[height=.53\textheight]{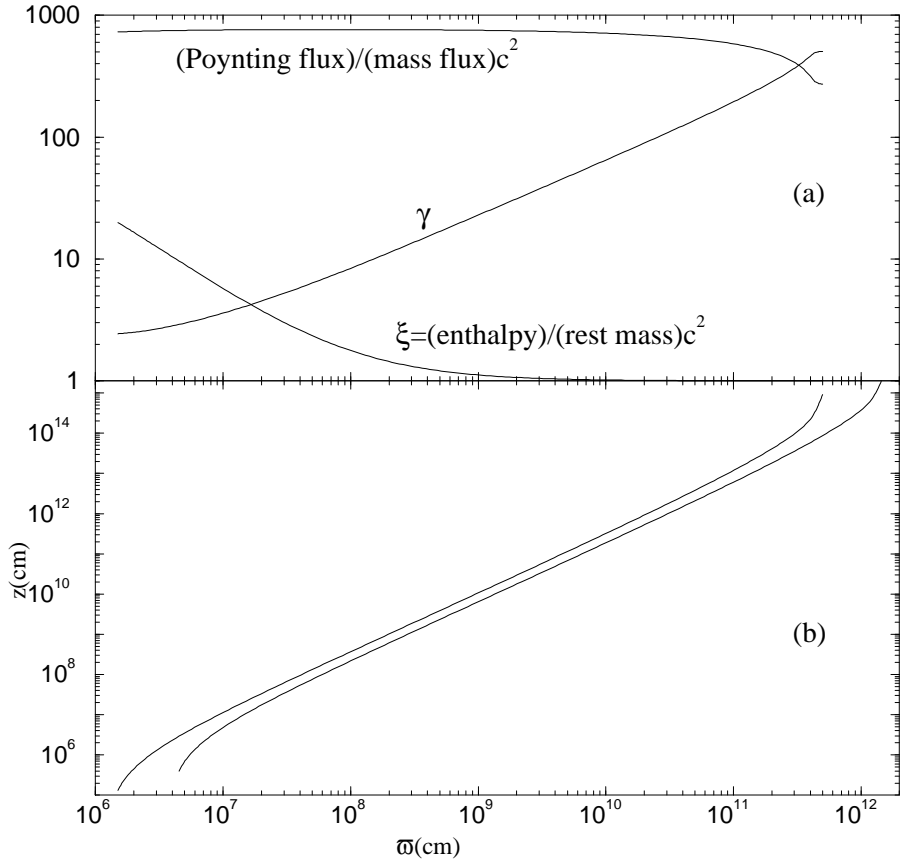}
\caption{Same as Fig. \ref{fig1}, but for a super-Alfv\'enic solution.
Here $\gamma \propto \varpi^{0.46}$, $z \propto \varpi^{1.48}$.
}\label{fig2}
\end{figure}

A representative super-Alfv\'enic solution, corresponding to
a base magnetic field 
($B_\phi \sim 10^{14}\ {\rm G}$, $B_p
\sim 10^{-2} B_\phi$), is shown in
Fig. \ref{fig2}. The super-Alfv\'enic solutions are
distinguished from the trans-Alfv\'enic ones
in two main respects: (1) During the initial thermal-acceleration
phase, some of the internal energy is transformed into
electromagnetic energy even as another part is used to increase
$V_p$. (2) During the subsequent magnetic-acceleration phase, the rate
of increase of the Lorentz factor with $z$
can be significantly lower than in the trans-Alfv\'enic case; the rate of
increase of the jet radius with $z$ is correspondingly higher. 
The overall magnetic-to-kinetic energy conversion efficiency is higher. 
See \citep{VK03b} for further details and
analytic scaling relations.

Another potentially important aspect of super-Alfv\'enic outflows,
namely, the possibility that their initial composition is highly
neutron-rich, could significantly alleviate the GRB baryon-loading problem.
In \citep{VPK03} (see also K\"onigl's contribution) it is shown that,
in contrast to the purely hydrodynamic case, the neutrons can
decouple at a Lorentz factor that is over an order of magnitude smaller than
$\gamma_{\infty}$ for the protons.

\end{document}